\documentclass[%
 aip,
rsi,%
 amsmath,amssymb,
 reprint,%
preprintnumbers,
]{revtex4-1}

\usepackage{graphicx}% Include figure files
\usepackage{dcolumn}% Align table columns on decimal point
\usepackage{bm}% bold math
\usepackage{bbold}
\begin{document}

\preprint{PI/UAN-2015-590FT}

\title[Finding Out the Gauge Field Strength Tensor]{A New Pedagogical Way of Finding Out the Gauge Field Strength Tensor\\in Abelian and Non-Abelian Local Gauge Field Theories} %\footnote{Error!}}% Force line breaks with \\
\thanks{To the memory of my friend and remarkable Colombian scientist Rodolfo A. D\'{\i}az.}

\author{Yeinzon Rodr\'{\i}guez}
%\altaffiliation[Simons Associate at ]{The Abdus Salam International Centre for Theoretical Physics, Trieste, Italy.} %Lines break automatically or can be forced with \\
%\author{B. Author}%
 \email{yeinzon.rodriguez@uan.edu.co.}
\affiliation{ 
Centro de Investigaciones en Ciencias B\'asicas y Aplicadas, Universidad Antonio Nari\~no, \\ Cra 3 Este \# 47A-15, Bogot\'a D.C. 110231, Colombia %\\This line break forced with \textbackslash\textbackslash
}%

%\author{C. Author}
% \homepage{http://www.Second.institution.edu/~Charlie.Author.}
\affiliation{
Escuela de F\'{\i}sica, Universidad Industrial de Santander, \\ Ciudad Universitaria, Bucaramanga 680002, Colombia %\\This line break forced% with \\
}%

\affiliation{Simons Associate at The Abdus Salam International Centre for Theoretical Physics, \\ Strada Costiera 11, I-34151, Trieste, Italy}

\date{\today}% It is always \today, today,
             %  but any date may be explicitly specified

\begin{abstract}
The gauge field strength tensor $F_{\mu \nu}$ in Abelian and non-Abelian local gauge field theories is a key object in the construction of the Lagrangian since it provides the kinetic term(s) of the gauge field(s) $A_\mu$.  When introducing this object, most of textbooks employ as a tool the commutator of the gauge covariant derivatives $D_\mu \psi$ of a fermion field $\psi$: $F_{\mu \nu} \psi = (i/g)[D_\mu,D_\nu]\psi$.  I argue that such a construction, although completely correct and valid, is not pedagogical enough for many students that approach the gauge field theories for the first time.  Another construction, based on the object $D_\mu A_\nu$: $F_{\mu \nu} = D_{[\mu} A_{\nu]}$, which avoids the introduction of additional and, for the case in consideration, spurious degrees of freedom such as the fermion one, simpler, more pedagogical in many cases, and suitable for first-time students, is presented.
%Valid PACS numbers may be entered using the \verb+\pacs{#1}+ command.
\end{abstract}

\pacs{11.15.-q}% PACS, the Physics and Astronomy
                             % Classification Scheme.
\keywords{Gauge field theories}%Use showkeys class option if keyword
                              %display desired
\maketitle

%\begin{quotation}
%The ``lead paragraph'' is encapsulated with the \LaTeX\ 
%\verb+quotation+ environment and is formatted as a single paragraph before the first section heading. 
%(The \verb+quotation+ environment reverts to its usual meaning after the first sectioning command.) 
%Note that numbered references are allowed in the lead paragraph.
% The lead paragraph will only be found in an article being prepared for the journal \textit{Chaos}.
%\end{quotation}

\section{Introduction}

Local gauge field theories have demonstrated their enormous power at describing the known fundamental interactions\cite{weinberg,hey,peskin,quigg,grif,lewis,mohapatra,gross,lyth,uti,kib,ald}.  Anybody interested in addressing the elementary particle physics, at least from the theoretical point of view, must learn the principles of the gauge field theories.  In that respect, textbooks do a good job taking the reader from the Abelian to the non-Abelian gauge field theories:  they start from the issue of the local gauge non-invariance of the Dirac's Lagrangian, continue showing the necessity of introducing a() gauge field(s) (which is(are) a() vector(s) in spacetime) with suitable transformation properties in order to make the Dirac's Lagrangian be gauge invariant, and finally introduce the gauge field strength tensor so that the gauge field(s) may have a() kinetic term(s) in the Lagrangian. The detailed procedure followed by most of textbooks, see e.g. Refs. \onlinecite{peskin,weinberg,quigg,hey,lewis}, which I will review later, involves the introduction of the commutator of the gauge covariant derivatives of a fermion field as a tool to arrive to the gauge field strength tensor and the way it transforms.  Despite the fact that this is the standard way of presenting things in textbooks and in the classroom, I argue that such a construction is in itself not pedagogical enough for many students that approach for the first time these subjects (for an alternative and very good pedagogical construction, see Ref. \onlinecite{uti}).  As I will show, such claimed lack of enough pedagogy lies on the fact that the actual issue of the gauge invariance of a term in the Lagrangian built {\it only} from first-order derivatives {\it of the gauge field(s)} is not confronted directly;  instead, the procedure lies on the introduction of the commutator of gauge covariant derivatives {\it of a fermion field} on whose result, curiously, we want to get rid of the fermion field itself.
%whose result, curiously, we do not want it to become an operation on partial derivatives of the fermion field itself.
%involve derivative operations on the fermion field itself in.  
Only after making use of this tool, the way the resultant object (which turns out to be the gauge field strength tensor operating on the fermion field) transforms is investigated taking advantage of the transformation rules of the fermion field and its gauge covariant derivatives.  Such knowledge finally allows us to build a gauge-invariant term in the Lagrangian, that contains the kinetic term(s) for the gauge field(s), from the gauge field strength tensor.  The procedure is clean and, at the end, satisfactory, especially when a deep geometrical meaning can be assigned to the commutator in terms of a round trip by parallel transport in exactly the same way as it is done in General Relativity to define the Riemann-Christoffel tensor\cite{weinberg,lewis}; however, for first-time students, especially for those that do not have an acquaintance in differential geometry, the philosophy behind the procedure may look strange, if not quite obscure, and the assimilation of the procedure relies more on the fact that it works and that it is presented in textbooks.

With the purpose of offering to students, especially those addressing the local gauge field theories for the first time, a simpler and perhaps more pedagogical construction of the gauge field strength tensor, I will attack directly the heart of the problem:  the gauge invariance of a term in the Lagrangian which comes from an object built {\it only} from first-order derivatives of the gauge field(s) {\it without the intrusion of other fields}, in particular fermion ones.  The philosophy behind this procedure is completely transparent and I hope students will take advantage of it when trying to understand where the gauge field strength tensor really comes from.

The layout of this paper is the following:  in Section \ref{abe}, I will present the Abelian local gauge field theories; Subsection \ref{pre} discusses the preliminaries, i.e., the issue of the local gauge non-invariance of the Dirac's Lagrangian and its resolution by the introduction of the gauge covariant derivative; Subsection \ref{fwabe} presents the standard way of finding out the gauge field strength tensor by means of the commutator of gauge covariant derivatives of a fermion field - this subsection is just a review devoted to people already familiarized with the standard procedure:  {\it first-time students should skip this subsection for better pedagogical results};  Subsection \ref{swabe} introduces the new way, simpler and perhaps more pedagogical, to find out the gauge field strength tensor:  {\it first-time students must definitely read this subsection}.  Section \ref{nonabe} deals with the non-Abelian local gauge field theories following the same strategy as in the Abelian case; therefore {\it first-time students should skip Subsection \ref{1nonabe} but must definitely read Subsection \ref{2nonabe}}.  Finally, I conclude in Section \ref{con}.  The Appendix complements the main text by briefly presenting the main ideas and calculations in explicit representations of the Lie algebra.

\section{\label{abe}Abelian local gauge field theories}

\subsection{\label{pre}Preliminaries}

In the Abelian local gauge field theories, the transformations over a fermion field commute.  I will consider the transformations under the $U(1)$ gauge group.  The fermion field $\psi$ transforms as
\begin{equation}
\psi' = e^{ig \epsilon(\vec{x})} \psi \,, \label{fundrep}
\end{equation}
where a prime denotes the transformed quantity, $g$ is the coupling constant, and $\epsilon(\vec{x})$ is a scalar quantity that denotes the amount of the transformation which, in turn, depends on the space location.

The idea is that the whole Lagrangian describing a fundamental interaction be invariant under transformations of some selected groups.  The first piece in the whole Lagrangian is the Dirac's Lagrangian, the one that describes the mass and kinetic properties of a fermion field:
\begin{equation}
\mathcal{L}_D = \overline{\psi}(i \gamma^\mu \partial_\mu - m) \psi \,.
\end{equation}
Here, $\overline{\psi}$ is the conjugate spinor associated to the fermion field $\psi$, $\gamma^\mu$ makes reference to the Dirac's matrices, and $m$ is the mass of the fermion field.
If the transformations were global, i.e. if $\epsilon$ did not depend on $\vec{x}$, $\partial_\mu \psi$ would transform as $\psi$ does, making the Dirac's Lagrangian be gauge invariant but, since I am considering local gauge transformations, $\partial_\mu \psi$ transforms as
\begin{equation}
(\partial_\mu \psi)' = ig (\partial_\mu \epsilon(\vec{x})) e^{ig \epsilon(\vec{x})} \psi + e^{ig \epsilon(\vec{x})} \partial_\mu \psi \,,
\end{equation}
which, of course, ruins the gauge invariance of $\mathcal{L}_D$.

In order to get rid of the annoying $\partial_\mu \epsilon(\vec{x})$ factor in the latter expression, we are required to introduce a gauge field $A_\mu$ that, together with $\partial_\mu$, and operating on $\psi$, replaces $\partial_\mu \psi$ itself;  such a construct is called the gauge covariant derivative of the fermion field, which means an object, similar to an ordinary derivative, that does transform as the fermion field:
\begin{equation}
D_\mu \psi = \partial_\mu \psi - ig A_\mu \psi \,. \label{covder}
\end{equation}
Thus, the new Dirac's Lagrangian is
\begin{equation}
\mathcal{L}_D = \overline{\psi}(i \gamma^\mu D_\mu - m) \psi \,.
\end{equation}
The gauge field $A_\mu$ must comply with a suitable transformation rule so that the $\partial_\mu \epsilon(\vec{x})$ factor disappears and $\mathcal{L}_D$ becomes gauge invariant: since we require
\begin{equation}
(D_\mu \psi)' = e^{ig \epsilon(\vec{x})} D_\mu \psi \,,
\end{equation}
$A_\mu$ must transform as
\begin{equation}
A_\mu' = A_\mu + \partial_\mu \epsilon(\vec{x}) \,.
\end{equation}
The interesting thing about $A_\mu$ is that it introduces interactions between the fermion field and its antiparticle field, $A_\mu$ being the field messenger of the fundamental interaction described by the selected group, in this case $U(1)$.

\subsection{\label{fwabe}Finding out the gauge field strength tensor: the standard way}

With the preliminaries above, the whole gauge-invariant Lagrangian can then be built from $\psi$ and $D_\mu \psi$, but not from $A_\mu$ (except when appearing in gauge covariant derivatives).  However, we need to introduce a gauge-invariant term in the Lagrangian that contains a kinetic term for the gauge field, i.e., a free-particle term quadratic in $\partial_\mu A_\nu$, so we have to consider terms of this form.

Where can we find a partial derivative of the gauge field?  We can look at Eq. (\ref{covder}) and say:  well, $A_\mu$ is there, so if we need a term of the form $\partial_\mu A_\nu$, the best we can do is to take the derivative of $D_\mu \psi$, a covariant derivative indeed to keep the good transformation properties.  But there is a serious issue here:  we want terms of the form $\partial_\mu A_\nu$, involving {\it only the gauge field}, at most operating on a fermion field, i.e. $(\partial_\mu A_\nu) \psi$ (just to get rid of the fermion field later), with no reference at all to derivatives of the fermion field.  At first sight, this seems to be impossible:
\begin{eqnarray}
D_\mu D_\nu \psi &=& (\partial_\mu - ig A_\mu)(\partial_\nu \psi - ig A_\nu \psi) \nonumber \\
&=& \partial_\mu \partial_\nu \psi - ig (\partial_\mu A_\nu) \psi \nonumber \\
&& -ig A_\nu (\partial_\mu \psi) -ig A_\mu (\partial_\nu \psi) - g^2 A_\mu A_\nu \psi \,. \label{introducingphi}
\end{eqnarray}
The only valuable term in the expression above is the second term in the second line;  the others can be considered as noise, especially the ones involving derivatives of the fermion field.  What can we do?: the first thing is to try avoiding the second-order partial derivative of $\psi$; a clever way to do it is antisymmetrizing the second-order covariant derivative:
\begin{eqnarray}
D_\mu D_\nu \psi - D_\nu D_\mu \psi &=& \partial_\mu \partial_\nu \psi - ig (\partial_\mu A_\nu) \psi -ig A_\nu (\partial_\mu \psi) \nonumber \\ && -ig A_\mu (\partial_\nu \psi) - g^2 A_\mu A_\nu \psi \nonumber \\
&& -\partial_\nu \partial_\mu \psi + ig (\partial_\nu A_\mu) \psi + ig A_\mu (\partial_\nu \psi) \nonumber \\
&& + ig A_\nu (\partial_\mu \psi) + g^2 A_\nu A_\mu \psi \,,
\end{eqnarray}
i.e.
\begin{equation}
[D_\mu,D_\nu]\psi = -ig (\partial_\mu A_\nu - \partial_\nu A_\mu) \psi \,. \label{maineq}
\end{equation}
Surprisingly, the antisymmetrization got rid not only of the second-order partial derivative of $\psi$ but also of all the other noisy terms, leaving only the valuable terms!  We will call $\partial_\mu A_\nu - \partial_\nu A_\mu$ the gauge field strength tensor $F_{\mu \nu}$:
\begin{equation}
F_{\mu \nu} = \partial_\mu A_\nu - \partial_\nu A_\mu \,,
\end{equation}
i.e.,
\begin{equation}
F_{\mu \nu} = \frac{i}{g} [D_\mu,D_\nu] \,. \label{bigeq1}
\end{equation}

Since we need to build a gauge-invariant kinetic term of the gauge field in the Lagrangian, and this term will be supposed to be built from $F_{\mu \nu}$, we need to know the way it transforms.  In order to do it, we will take advantage of the fact that $[D_\mu,D_\nu]\psi$ transforms as $\psi$.  Thus,
\begin{eqnarray}
([D_\mu,D_\nu]\psi)' &=& e^{ig \epsilon(\vec{x})} [D_\mu,D_\nu]\psi \nonumber \\
&=& e^{ig \epsilon(\vec{x})} (-ig F_{\mu \nu} \psi) \,, \label{ale1}
\end{eqnarray}
but, on the other hand,
\begin{eqnarray}
(-ig F_{\mu \nu} \psi)' &=& -ig F_{\mu \nu}' \psi' \nonumber \\
&=& e^{ig \epsilon(\vec{x})} (-ig F_{\mu \nu}' \psi) \,, \label{ale2}
\end{eqnarray}
therefore, comparing Eqs. (\ref{ale1}) and (\ref{ale2}), we conclude that the gauge field strength tensor $F_{\mu \nu}$ is gauge invariant.  This is marvellous since the Lorentz-invariant Lagrangian $\mathcal{L}_{K-A}$:
\begin{equation}
\mathcal{L}_{K-A} = -\frac{1}{4} F_{\mu \nu} F^{\mu \nu} \,,
\end{equation}
built as a term that contains a free-particle term quadratic in $\partial_\mu A_\nu$,
is gauge invariant.  This is the kinetic term in the Lagrangian associated to the gauge field.

The main criticism to this whole procedure is the non confrontation of the actual issue:  finding out a gauge-invariant term in the Lagrangian built {\it only} from derivatives of the gauge field {\it without introducing}, ever, other fields.  Instead, a fermion field is introduced in Eq. (\ref{introducingphi}) just for us to get rid of it when going from Eq. (\ref{maineq}) to Eq. (\ref{bigeq1}) (of course, taking into account Eqs. (\ref{ale1}) and (\ref{ale2})).  
%The procedure is not at all difficult to tackle with, but 
Although 
%it is important to admit that 
it is not so absurd not to consider the gauge field independently of the fermion one since the former appears from the necessity of building a covariant derivative of the latter, the unnecessary introduction of the fermion field in order to find out the gauge field strength tensor, unnecessary in the sense that the fermion field is just a dispensable artefact to arrive to the desired result, may make the procedure look strange, if not quite obscure, for non-familiarized readers.  It is worthwhile noticing, however, that the student familiarized with differential geometry will find this procedure more illuminating than the one I propose, in the next subsection, as being perhaps more pedagogical;  the deep geometrical meaning of the commutator of covariant derivatives as well as its strong analogy with the procedures followed in General Relativity to arrive to the mathematical objects that describe the space-time curvature\cite{weinberg,lewis} are, without any doubt, robust arguments to choose the standard way described in this subsection as the most pedagogical and illuminating.  Notwithstanding, it is also worthwhile remembering that neophyte students usually do not master such knowledge.

\subsection{\label{swabe}Finding out the gauge field strength tensor: a simpler and perhaps more pedagogical way}

The preliminaries presented in Subsection \ref{pre} lead to conclude that in the construction of the whole gauge-invariant Lagrangian, only $\psi$ and $D_\mu \psi$ can be used; the only way $A_\mu$ can be introduced is via the gauge covariant derivatives.  Despite of this fact, a free-particle term quadratic in $\partial_\mu A_\nu$ that represents the kinetic term for the gauge field must be contained in some gauge-invariant term in the Lagrangian. We have, then, to consider terms of this form.
%order to  However, we need to introduce a gauge-invariant term in the Lagrangian that contains a kinetic term for the gauge field, i.e., a free-particle term quadratic in $\partial_\mu A_\nu$, so we have to consider terms of this form.

Well, since the protagonist object here is $\partial_\mu A_\nu$, let's find out {\it directly} how it transforms:
\begin{equation}
(\partial_\mu A_\nu)' = \partial_\mu (A_\nu + \partial_\nu \epsilon(\vec{x})) = \partial_\mu A_\nu + \partial_\mu \partial_\nu \epsilon(\vec{x}) \,.
\end{equation}
As in the Dirac's Lagrangian, partial derivatives of $\epsilon(\vec{x})$ are annoying terms that we must get rid of. How do we do it? The antisymmetrization of $\partial_\mu A_\nu$ seems to be the answer:
\begin{eqnarray}
(\partial_\mu A_\nu - \partial_\nu A_\mu)' &=&  \partial_\mu A_\nu + \partial_\mu \partial_\nu \epsilon(\vec{x}) - \partial_\nu A_\mu - \partial_\nu \partial_\mu \epsilon(\vec{x}) \nonumber \\
&=& \partial_\mu A_\nu - \partial_\nu A_\mu \,.
\end{eqnarray}
This is fantastic!:  we have found a gauge-invariant object built only from derivatives of the gauge field.  We will call it the gauge field strength tensor $F_{\mu \nu}$:
\begin{equation}
F_{\mu \nu} = \partial_\mu A_\nu - \partial_\nu A_\mu \,.
\end{equation}

Thus, since the objective is to find out a gauge-invariant term in the Lagrangian that contains a kinetic term for the gauge field, the latter being, as said before, a free-particle term which is quadratic in $\partial_\mu A_\nu$, the simplest thing to do is to have a Lorentz-invariant product of $F_{\mu \nu}$ with itself:
\begin{equation}
\mathcal{L}_{K-A} = -\frac{1}{4} F_{\mu \nu} F^{\mu \nu} \,.
\end{equation}

The reader, at this point, may think that there is nothing novel in the approach I am presenting in this subsection because, up to this point, everything in this subsection is already discussed in textbooks.  The latter is true but its direct generalization to non-Abelian gauge field theories does not appear in textbooks and, as the reader will notice, all the power of this procedure will reveal when discussing such theories in Subsection \ref{2nonabe}.  Indeed, the following, which appears as a curiosity in the Abelian case, will turn out to be fundamental in the non-Abelian case: the gauge field strength tensor can also be written as
\begin{eqnarray}
F_{\mu \nu} &=& \partial_\mu A_\nu - \partial_\nu A_\mu \nonumber \\
&=& \partial_\mu A_\nu -ig A_\mu A_\nu - \partial_\nu A_\mu + ig A_\nu A_\mu \nonumber \\
&=& D_\mu A_\nu - D_\nu A_\mu \nonumber \\
&=& D_{[\mu} A_{\nu]} \,, \label{magic}
\end{eqnarray}
where the symbol $D_\mu$ is the same one as that defined in Eq. (\ref{covder}).  It is very important to stress that  $D_\mu A_\nu$ {\it is not at all} a covariant derivative since such an object does not transform as a field belonging to a given representation of the $U(1)$ gauge group (e.g., the fundamental representation, see Eq. (\ref{fundrep}), or the adjoint one):
\begin{eqnarray}
(D_\mu A_\nu)' &=& \partial_\mu A_\nu' - ig A_\mu' A_\nu' \nonumber \\
&=& \partial_\mu (A_\nu + \partial_\nu \epsilon(\vec{x})) \nonumber \\
&& - ig (A_\mu + \partial_\mu \epsilon(\vec{x})) (A_\nu + \partial_\nu \epsilon(\vec{x})) \nonumber \\
&=& \partial_\mu A_\nu + \partial_\mu \partial_\nu \epsilon(\vec{x}) -ig [A_\mu A_\nu + A_\mu (\partial_\nu \epsilon(\vec{x})) \nonumber \\
&& + (\partial_\mu \epsilon(\vec{x})) A_\nu + \partial_\mu \epsilon(\vec{x}) \partial_\nu \epsilon(\vec{x})] \,.
\end{eqnarray}

The expression $F_{\mu \nu} = D_{[\mu} A_{\nu]}$ should make us think that, maybe, the construction $D_{[\mu} A_{\nu]}$ is a good candidate to define the gauge field strength tensor in the non-Abelian case too.  The next section will show that this is indeed the case, but, before going there, it is important to make an annotation. The expression $F_{\mu \nu} = D_{[\mu} A_{\nu]}$ had already been employed in Refs. \onlinecite{gross,lyth} and, very marginally, in Ref. \onlinecite{uti}, the first two references in the context of non-Abelian local gauge field theories, without giving a dedicated reasoning about why $F_{\mu \nu}$ is constructed the way it is, which, of course, is not sufficiently valuable from the pedagogical point of view,  and the third reference when implementing the gauge-field approach to gravity. 

\section{\label{nonabe}Non-Abelian local gauge field theories}

\subsection{\label{prenon}Preliminaries}

In the non-Abelian local gauge field theories, the transformations over a fermion field do not commute.  I will consider the transformations under the $SU(N)$ gauge group.  The fermion field $\psi$ is an $N$-dimensional spinor that transforms as
\begin{equation}
\psi' = e^{ig \vec{\epsilon}(\vec{x}) \cdot \vec{T}} \psi \,,
\end{equation}
where $\vec{\epsilon}(\vec{x})$ is an $(N^2 - 1)$-dimensional vector that denotes the amount of the transformation which, in turn, depends on the space location, and $\vec{T}$ is the ``vector'' built with the $N^2 - 1$ generators of the $SU(N)$ group that can be represented as $N \times N$ matrices that satisfy the following simple Lie algebra:
\begin{equation}
[T_a,T_b] = i f^c_{\;\;ab} T_c \,. \label{generators}
\end{equation}
In the latter expression, $f^c_{\;\;ab}$ are the structure constants of the group that turn out to be totally antisymmetric, and $a,b,c$ run from 1 to $N^2 - 1$.

The Dirac's Lagrangian in this case is
\begin{equation}
\mathcal{L}_D = \overline{\psi}[i \gamma^\mu (\partial_\mu \mathbb{1}) - m \mathbb{1}] \psi \,,
\end{equation}
where $\mathbb{1}$ is the $N \times N$ unit matrix.
If the transformations were global, i.e. if $\vec{\epsilon}$ did not depend on $\vec{x}$, $(\partial_\mu \mathbb{1}) \psi$ would transform as $\psi$ does, making the Dirac's Lagrangian be gauge invariant but, since I am considering local gauge transformations, $(\partial_\mu \mathbb{1}) \psi$ transforms as
\begin{equation}
[(\partial_\mu \mathbb{1}) \psi]' = (\partial_\mu e^{ig \vec{\epsilon}(\vec{x}) \cdot \vec{T}}) \psi + e^{ig \vec{\epsilon}(\vec{x}) \cdot \vec{T}} (\partial_\mu \mathbb{1}) \psi \,,
\end{equation}
which, of course, ruins the gauge invariance of $\mathcal{L}_D$.

In order to get rid of the annoying $\partial_\mu e^{ig \vec{\epsilon}(\vec{x}) \cdot \vec{T}}$ factor in the latter expression, we are required to introduce $N^2 - 1$ gauge fields $A^a_\mu$ that are grouped into a single matrix by using the group generators $T_a$:
\begin{equation}
A_\mu = A^a_\mu T_a \,.
\end{equation}
Such a matrix-gauge field, together with $\partial_\mu \mathbb{1}$, and operating on $\psi$, defines the covariant derivative of the fermion field and replaces $(\partial_\mu \mathbb{1}) \psi$:
\begin{equation}
D_\mu \psi = (\partial_\mu \mathbb{1}) \psi - ig A_\mu \psi \,.  \label{covdernona}
\end{equation}
As such, the covariant derivative of the fermion field transforms as the fermion field itself:
\begin{equation}
(D_\mu \psi)' = e^{ig \vec{\epsilon}(\vec{x}) \cdot \vec{T}} D_\mu \psi \,.
\end{equation}
The new Dirac's Lagrangian is then
\begin{equation}
\mathcal{L}_D = \overline{\psi}(i \gamma^\mu D_\mu - m \mathbb{1}) \psi \,,
\end{equation}
which is gauge invariant as long as the matrix-gauge field $A_\mu$ complies with a suitable transformation rule so that the $\partial_\mu e^{ig \vec{\epsilon}(\vec{x}) \cdot \vec{T}}$ factor disappears:
\begin{equation}
A_\mu' = e^{ig \vec{\epsilon}(\vec{x}) \cdot \vec{T}} A_\mu e^{-ig \vec{\epsilon}(\vec{x}) \cdot \vec{T}} - \frac{i}{g} (\partial_\mu e^{ig \vec{\epsilon}(\vec{x}) \cdot \vec{T}}) e^{-ig \vec{\epsilon}(\vec{x}) \cdot \vec{T}} \,. \label{nontransf}
\end{equation}
Similar to $A_\mu$ in the Abelian case, the non-Abelian gauge fields $A^a_\mu$ introduce interactions among different fermion fields and are the field messengers of the fundamental interaction described by the $SU(N)$ group.

\subsection{\label{1nonabe}Finding out the gauge field strength tensor: the standard way}

As with the local Abelian gauge field theories, the whole gauge-invariant Lagrangian can then be built from $\psi$ and $D_\mu \psi$, but not from $A_\mu$ (except when appearing in gauge covariant derivatives).  The missing pieces in the Lagrangian are the kinetic terms for the $N^2 - 1$ gauge fields which, as we already know, must be free-particle terms quadratic in $\partial_\mu A^a_\nu$, all of them coming from a gauge-invariant term.  We then have to consider terms of this form.

The same question asked in the Abelian case applies here:  where can we find a partial derivative of the gauge fields?  Looking at Eq. (\ref{covdernona}) we can conclude that, since $A_\mu$ is there, we must take the derivative of $D_\mu \psi$ in order to get $\partial_\mu A_\nu$ and, therefore, the desired $\partial_\mu A^a_\nu$.  Such a derivative should be a covariant one in order to keep the good transformation properties.  However, we want terms of the form $\partial_\mu A_\nu$, involving {\it only the matrix-gauge field}, at most operating on a fermion field, i.e. $(\partial_\mu A_\nu) \psi$ (just to get rid of the fermion field later), with no reference at all to derivatives of the fermion field.  Preliminary calculations show this to seem to be impossible:
\begin{eqnarray}
D_\mu D_\nu \psi &=& [(\partial_\mu \mathbb{1}) - ig A_\mu][(\partial_\nu \mathbb{1}) \psi - ig A_\nu \psi] \nonumber \\
&=& (\partial_\mu \partial_\nu \mathbb{1}) \psi - ig (\partial_\mu A_\nu) \psi \nonumber \\
&& -ig A_\nu [(\partial_\mu \mathbb{1}) \psi] -ig A_\mu [(\partial_\nu \mathbb{1}) \psi] - g^2 A_\mu A_\nu \psi \,. \nonumber \\
&& \label{doublenona}
\end{eqnarray}
Since the only valuable term in the expression above is the second term in the second line, the other terms being considered possibly as noise, especially the ones involving derivatives of the fermion field, the first thing to do is to try avoiding the second-order partial derivative of $\psi$. This can be realized by antisymmetrizing the second-order covariant derivative:
\begin{eqnarray}
D_\mu D_\nu \psi - D_\nu D_\mu \psi &=& (\partial_\mu \partial_\nu \mathbb{1}) \psi - ig (\partial_\mu A_\nu) \psi \nonumber \\
&& -ig A_\nu [(\partial_\mu \mathbb{1}) \psi] \nonumber -ig A_\mu [(\partial_\nu \mathbb{1}) \psi] \nonumber \\
&& - g^2 A_\mu A_\nu \psi \nonumber \\
&& - (\partial_\nu \partial_\mu \mathbb{1}) \psi + ig (\partial_\nu A_\mu) \psi \nonumber \\
&& + ig A_\mu [(\partial_\nu \mathbb{1}) \psi] + ig A_\nu [(\partial_\mu \mathbb{1}) \psi] \nonumber \\
&& + g^2 A_\nu A_\mu \psi \,,
\end{eqnarray}
i.e.,
\begin{equation}
[D_\mu,D_\nu]\psi = -ig (\partial_\mu A_\nu - \partial_\nu A_\mu -ig [A_\mu,A_\nu]) \psi \,, \label{conmutenona}
\end{equation}
where we notice that, since the generators of the group do not commute, see Eq. (\ref{generators}), the matrix-gauge fields $A_\mu$ and $A_\nu$ do not commute either.
Similarly to the Abelian case, the antisymmetrization got rid, surprisingly, not only of the second-order partial derivative of $\psi$ but also of the terms involving first-order derivatives of $\psi$.  We will call $\partial_\mu A_\nu - \partial_\nu A_\mu -ig [A_\mu,A_\nu]$ the matrix-gauge field strength tensor $F_{\mu \nu}$:
\begin{equation}
F_{\mu \nu} = \partial_\mu A_\nu - \partial_\nu A_\mu -ig [A_\mu,A_\nu] \,, \label{strtennona}
\end{equation}
i.e.,
\begin{equation}
F_{\mu \nu} = \frac{i}{g} [D_\mu,D_\nu] \,. \label{bigeq2}
\end{equation}

The construction of the kinetic terms of the gauge fields from the matrix-gauge field strength tensor requires to know how the latter transforms. To know it, we will take advantage of the fact that $[D_\mu,D_\nu]\psi$ transforms as $\psi$.  Thus,
\begin{eqnarray}
([D_\mu,D_\nu]\psi)' &=& e^{ig \vec{\epsilon}(\vec{x}) \cdot \vec{T}} [D_\mu,D_\nu]\psi \nonumber \\
&=& e^{ig \vec{\epsilon}(\vec{x}) \cdot \vec{T}} (-ig F_{\mu \nu} \psi) \,, \label{alenon1}
\end{eqnarray}
but, on the other hand,
\begin{eqnarray}
(-ig F_{\mu \nu} \psi)' &=& -ig F_{\mu \nu}' \psi' \nonumber \\
&=& -ig F_{\mu \nu}' (e^{ig \vec{\epsilon}(\vec{x}) \cdot \vec{T}} \psi) \,, \label{alenon2}
\end{eqnarray}
therefore, comparing Eqs. (\ref{alenon1}) and (\ref{alenon2}), we conclude that the matrix-gauge field strength tensor $F_{\mu \nu}$ transforms in the adjoint representation\cite{weinberg} of the $SU(N)$ group:
\begin{equation}
F_{\mu \nu}' = e^{ig \vec{\epsilon}(\vec{x}) \cdot \vec{T}} \ F_{\mu \nu} \ e^{-ig \vec{\epsilon}(\vec{x}) \cdot \vec{T}} \,.
\end{equation}
It is in this sense that the gauge field strength tensor {\it can be considered} as the covariant derivative of the gauge field itself.
The situation here is quite different to that in the Abelian case since the matrix-gauge field strength tensor is not gauge invariant. So, how can we build a gauge-invariant quantity from it?  The answer relies on the fact that the searched gauge-invariant term in the Lagrangian is a scalar whereas the matrix-gauge field strength tensor is, of course, a matrix, so we have to get a scalar from a matrix.  The usual scalars we can get from a matrix are its determinant and its trace.  Indeed, if we take the trace of $F_{\mu \nu}'$, we can see that it is gauge invariant:
\begin{eqnarray}
Tr(F_{\mu \nu}') &=& Tr(e^{ig \vec{\epsilon}(\vec{x}) \cdot \vec{T}} \ F_{\mu \nu} \ e^{-ig \vec{\epsilon}(\vec{x}) \cdot \vec{T}}) \nonumber \\
&=& Tr(F_{\mu \nu} \ e^{-ig \vec{\epsilon}(\vec{x}) \cdot \vec{T}} e^{ig \vec{\epsilon}(\vec{x}) \cdot \vec{T}}) \nonumber \\
&=& Tr(F_{\mu \nu}) \,.
\end{eqnarray}
The same kind of reasoning applies to the Lorentz-invariant matrix object $-\frac{1}{2} F_{\mu \nu} F^{\mu \nu}$: it transforms as
\begin{equation}
\left(-\frac{1}{2} F_{\mu \nu} F^{\mu \nu}\right)' = e^{ig \vec{\epsilon}(\vec{x}) \cdot \vec{T}} \ \left(-\frac{1}{2} F_{\mu \nu} F^{\mu \nu}\right) \ e^{-ig \vec{\epsilon}(\vec{x}) \cdot \vec{T}} \,,
\end{equation}
so that its trace is gauge invariant;  this is precisely the Lorentz-invariant Lagrangian we are looking for:
\begin{eqnarray}
\mathcal{L}_{K-A} = Tr \left(-\frac{1}{2} F_{\mu \nu} F^{\mu \nu}\right) &=& -\frac{1}{2} g_{ab} F^a_{\mu \nu} F^{b \mu \nu} \nonumber \\
&=& -\frac{1}{4} \delta_{ab} F^a_{\mu \nu} F^{b \mu \nu} \,,
\end{eqnarray}
where
\begin{equation}
Tr(T_a T_b) = g_{ab} \,,
\end{equation}
is the induced metric on the group\cite{weinberg} $\left(g_{ab} = \frac{\delta_{ab}}{2}\right)$
and
\begin{equation}
F^a_{\mu \nu} = \partial_\mu A^a_\nu - \partial_\nu A^a_\mu + g f^a_{\;\;bc} A^b_\mu A^c_\nu \,.
\end{equation}
This is a Lagrangian that contains free-particle terms quadratic in $\partial_\mu A^a_\nu$, i.e., the kinetic terms of the gauge fields, but that also contains self-interaction terms among the different gauge fields due to the $-ig[A_\mu,A_\nu]$ term in Eq. (\ref{strtennona}).  So, in contrast with the local Abelian gauge field theories, where the gauge field does not interact with itself, in the local non-Abelian gauge field theories, the gauge fields interact among themselves.

The main criticism to this whole procedure is the same as in the Abelian case, i.e., the non confrontation of the actual issue:  finding out a gauge-invariant term in the Lagrangian built {\it only} from derivatives of the gauge fields {\it without introducing}, ever, other fields.  As Eq. (\ref{doublenona}) shows, a fermion field is introduced and later it is abandoned when going from Eq. (\ref{conmutenona}) to Eq. (\ref{bigeq2}) (taking into account Eqs. (\ref{alenon1}) and (\ref{alenon2})).  
%The procedure is not at all difficult to tackle with, but the unnecessary introduction of the fermion field, unnecessary in the sense that the fermion field is just a dispensable artefact to arrive to the desired result, may make the procedure look strange, if not quite obscure, for non-familiarized readers. 
The situation is not any better in other textbooks such as those in Refs. \onlinecite{grif,mohapatra}:  they just notice that the expression $\partial_\mu A_\nu - \partial_\nu A_\mu$ is neither gauge invariant nor transforms in an adequate way (involving $\vec{\epsilon}(\vec{x})$ but not its derivatives), and introduce the term $-ig [A_\mu,A_\nu]$, essentially from nowhere, showing that its transformation counteracts the inadequate way $\partial_\mu A_\nu - \partial_\nu A_\mu$ transforms.

\subsection{\label{2nonabe}Finding out the gauge field strength tensor: a simpler and perhaps more pedagogical way}

The preliminaries presented in Subsection \ref{prenon} lead to conclude that in the construction of the whole gauge-invariant Lagrangian, only $\psi$ and $D_\mu \psi$ can be used; the only way $A_\mu$ can be introduced is via the gauge covariant derivatives.  Despite of this fact, free-particle terms quadratic in $\partial_\mu A^a_\nu$ that represent the kinetic terms for the gauge fields must arise from some gauge-invariant term in the Lagrangian. We have, then, to consider terms of the form $\partial_\mu A_\nu$:  this is the protagonist object here.

Let's then find out {\it directly} how $\partial_\mu A_\nu$ transforms:
\begin{eqnarray}
(\partial_\mu A_\nu)' &=& \partial_\mu \Big[e^{ig \vec{\epsilon}(\vec{x}) \cdot \vec{T}} A_\nu e^{-ig \vec{\epsilon}(\vec{x}) \cdot \vec{T}} \nonumber \\
&& - \frac{i}{g} (\partial_\nu e^{ig \vec{\epsilon}(\vec{x}) \cdot \vec{T}}) e^{-ig \vec{\epsilon}(\vec{x}) \cdot \vec{T}}\Big] \nonumber \\
&=& \left[\partial_\mu (e^{ig \vec{\epsilon}(\vec{x}) \cdot \vec{T}})\right] A_\nu e^{-ig \vec{\epsilon}(\vec{x}) \cdot \vec{T}} \nonumber \\
&& + e^{ig \vec{\epsilon}(\vec{x}) \cdot \vec{T}} (\partial_\mu A_\nu) e^{-ig \vec{\epsilon}(\vec{x}) \cdot \vec{T}} \nonumber \\
&& + e^{ig \vec{\epsilon}(\vec{x}) \cdot \vec{T}} A_\nu \left[\partial_\mu (e^{-ig \vec{\epsilon}(\vec{x}) \cdot \vec{T}})\right] \nonumber \\
&& - \frac{i}{g} \left[\partial_\mu \partial_\nu (e^{ig \vec{\epsilon}(\vec{x}) \cdot \vec{T}})\right] e^{-ig \vec{\epsilon}(\vec{x}) \cdot \vec{T}} \nonumber \\ 
&& - \frac{i}{g} \left[\partial_\nu (e^{ig \vec{\epsilon}(\vec{x}) \cdot \vec{T}})\right] \left[\partial_\mu (e^{-ig \vec{\epsilon}(\vec{x}) \cdot \vec{T}})\right] \,.
\end{eqnarray}
As in the Dirac's Lagrangian, partial derivatives of $\vec{\epsilon}(\vec{x})$ are annoying terms that we must get rid of, and we have to do it, initially, by removing the most annoying of all of them: the second-order derivatives. As always, the antisymmetrization of $\partial_\mu A_\nu$ seems to be the answer:
\begin{eqnarray}
(\partial_\mu A_\nu - \partial_\nu A_\mu)' &=&  \left[\partial_{[\mu} (e^{ig \vec{\epsilon}(\vec{x}) \cdot \vec{T}})\right] A_{\nu]} e^{-ig \vec{\epsilon}(\vec{x}) \cdot \vec{T}} \nonumber \\
&& + e^{ig \vec{\epsilon}(\vec{x}) \cdot \vec{T}} (\partial_\mu A_\nu -\partial_\nu A_\mu) e^{-ig \vec{\epsilon}(\vec{x}) \cdot \vec{T}} \nonumber \\
&& - e^{ig \vec{\epsilon}(\vec{x}) \cdot \vec{T}} A_{[\mu} \left[\partial_{\nu]} (e^{-ig \vec{\epsilon}(\vec{x}) \cdot \vec{T}})\right] \nonumber \\
&& + \frac{i}{g} \left[\partial_{[\mu} (e^{ig \vec{\epsilon}(\vec{x}) \cdot \vec{T}})\right] \left[\partial_{\nu]} (e^{-ig \vec{\epsilon}(\vec{x}) \cdot \vec{T}})\right] \,. \nonumber \\
&&
\end{eqnarray}

In contrast with the Abelian case, the object we have just found still contains (first-order) derivatives of $\vec{\epsilon}(\vec{x})$ and, as the reader can easily check from the latter expression, the only way to get rid of them seems to drop the idea that the theory is non-Abelian.  This, of course, does not make any sense.  However, if we are led again to the local Abelian gauge field theories, we are reminded of an interesting curiosity (see Eq. (\ref{magic})): the gauge field strength tensor can also be written as the antisymmetric object $D_{[\mu} A_{\nu]}$.
Thus, why don't we try building an object in local non-Abelian gauge field theories following the same recipe?:
\begin{eqnarray}
D_{[\mu} A_{\nu]} &=& [(\partial_\mu \mathbb{1}) - igA_\mu] A_\nu - [(\partial_\nu \mathbb{1}) - igA_\nu] A_\mu \nonumber \\
&=& \partial_\mu A_\nu - \partial_\nu A_\mu -ig [A_\mu,A_\nu] \,.
\end{eqnarray}
We will call this object the matrix-gauge field strength tensor:
\begin{equation}
F_{\mu \nu} = \partial_\mu A_\nu - \partial_\nu A_\mu -ig [A_\mu,A_\nu] \,. \label{matgau}
\end{equation}
If we want to build a gauge-invariant Lagrangian from it, we need to know how it transforms. By employing Eq. (\ref{nontransf}), and performing a quite length but straightforward algebra, the reader can conclude that
\begin{equation}
F_{\mu \nu}' = e^{ig \vec{\epsilon}(\vec{x}) \cdot \vec{T}} \ F_{\mu \nu} \ e^{-ig \vec{\epsilon}(\vec{x}) \cdot \vec{T}} \,. \label{jacob}
\end{equation}
Good point!:  no spatial derivatives of $\vec{\epsilon}(\vec{x})$ appear and $F_{\mu \nu}$ turns out to transform in the adjoint representation of the $SU(N)$ group.  $F_{\mu \nu}$ {\it can be considered}, therefore, as the covariant derivative of the gauge field itself.

The challenge now is to build a gauge-invariant term from the matrix-gauge field strength tensor.  But how can we do such a thing when $F_{\mu \nu}$ is a matrix whereas any term in the Lagrangian must be a scalar? Well, the usual scalars we can get from a matrix are its determinant and its trace.  Observing the transformation property in Eq. (\ref{jacob}), we can recognize that, indeed, the trace of the matrix-gauge field strength tensor is gauge invariant:
\begin{eqnarray}
Tr(F_{\mu \nu}') &=& Tr(e^{ig \vec{\epsilon}(\vec{x}) \cdot \vec{T}} \ F_{\mu \nu} \ e^{-ig \vec{\epsilon}(\vec{x}) \cdot \vec{T}}) \nonumber \\
&=& Tr(F_{\mu \nu} \ e^{-ig \vec{\epsilon}(\vec{x}) \cdot \vec{T}} e^{ig \vec{\epsilon}(\vec{x}) \cdot \vec{T}}) \nonumber \\
&=& Tr(F_{\mu \nu}) \,,
\end{eqnarray}
as well as it is the trace of the Lorentz-invariant matrix object $-\frac{1}{2} F_{\mu \nu} F^{\mu \nu}$:
%The same kind of reasoning applies to the Lorentz-invariant matrix object $-\frac{1}{2} F_{\mu \nu} F^{\mu \nu}$: it transforms as
\begin{eqnarray}
&& \left[Tr\left(-\frac{1}{2} F_{\mu \nu} F^{\mu \nu}\right)\right]' \nonumber \\
&=& Tr\left[e^{ig \vec{\epsilon}(\vec{x}) \cdot \vec{T}} \ \left(-\frac{1}{2} F_{\mu \nu} F^{\mu \nu}\right) \ e^{-ig \vec{\epsilon}(\vec{x}) \cdot \vec{T}}\right] \nonumber \\
&=& Tr\left(-\frac{1}{2} F_{\mu \nu} F^{\mu \nu}\right) \,.
\end{eqnarray}
Thus, we have found the Lorentz-invariant Lagrangian we were looking for:
\begin{eqnarray}
\mathcal{L}_{K-A} = Tr \left(-\frac{1}{2} F_{\mu \nu} F^{\mu \nu}\right) &=& -\frac{1}{2} g_{ab} F^a_{\mu \nu} F^{b \mu \nu} \nonumber \\
&=& -\frac{1}{4} \delta_{ab} F^a_{\mu \nu} F^{b \mu \nu} \,, \label{chav}
\end{eqnarray}
where
\begin{equation}
Tr(T_a T_b) = g_{ab} \,,
\end{equation}
is the induced metric on the group\cite{weinberg} $\left(g_{ab} = \frac{\delta_{ab}}{2}\right)$
and
\begin{equation}
F^a_{\mu \nu} = \partial_\mu A^a_\nu - \partial_\nu A^a_\mu + g f^a_{\;\;bc} A^b_\mu A^c_\nu \,.
\end{equation}
The kinetic terms of the gauge fields, i.e., the free-particle terms quadratic in $\partial_\mu A^a_\nu$, are contained in the Lagrangian in Eq. (\ref{chav}).  But there are more terms contained in such a Lagrangian:  because of the $-ig[A_\mu,A_\nu]$ term in Eq. (\ref{matgau}), there are now self-interaction terms among the different gauge fields in contrast with what happens in local Abelian gauge field theories.

\section{\label{con}Conclusions}

Local Abelian and non-Abelian gauge field theories are essential for anybody who wishes to understand the way the fundamental interactions are described nowadays.  Most of textbooks, e.g. Refs. \onlinecite{peskin,weinberg,quigg,hey,lewis}, follow a standard procedure to introduce such theories, finding out, specifically, the gauge field strength tensor from the commutator of covariant derivatives $D_\mu \psi$ of a fermion field $\psi$: $F_{\mu \nu} \psi = (i/g)[D_\mu,D_\nu]\psi$.  I have argued that such an approach is not pedagogical enough for many students approaching for the first time these subjects since the issue of getting a gauge-invariant Lagrangian, containing the kinetic term(s) of the gauge field(s), from first-order derivatives of the gauge field(s) {\it only}, is not confronted directly;  this applies particularly to students with no acquaintance in differential geometry as it is impossible, this way, to link the commutator of covariant derivatives with a round trip by parallel transport and observe the connection of this procedure with the one followed in General Relativity to find out the mathematical objects that describe the space-time curvature\cite{lewis,weinberg}.  In this paper, such a direct confrontation has been performed, finding positive results when comparing this procedure with the usual one.  In particular, I have presented a very simple and nice way of constructing the gauge field strength tensor:  this object is the antisymmetrized version of $D_\mu A_\nu$:  $F_{\mu \nu} = D_{[\mu} A_{\nu]}$.  Since such a construction can have an enormous pedagogical value for many first-time students, I expect this methodology can help them to grasp, in a quicker and more efficient way, the profound question of where the gauge field strength tensor really comes from.

\begin{acknowledgments}
This work was supported by COLCIENCIAS - ECOS NORD grant number RC 0899-2012 with the help of ICETEX, and by COLCIENCIAS grant numbers 110656933958 RC 0384-2013 and 123365843539 RC FP44842-081-2014.  My gratitude goes to Edison A. Montoya, Carlos M. Nieto, Erwan Allys, Patrick Peter, Rodolfo A. D\'{\i}az, Eduar A. Becerra, Stanley Deser, and David H. Lyth for their useful comments and criticisms.  Rodolfo A. D\'{\i}az, my friend and a very valuable Colombian scientist, passed away while I was producing one of the latest versions of this paper; this final document is dedicated to his memory.
\end{acknowledgments}

\appendix*

\section{}

\subsection{Preliminaries} \label{prel}

I will work in the framework of {\it local} gauge transformations that are part of a simple Lie group.  A fermion field $\psi_l$ will transform under an infinitesimal transformation as:
\begin{equation}
\delta \psi_l = i g \epsilon^a(\vec{x}) (T_a)_l^{\;\;m} \psi_m \,,
\end{equation}
where $\epsilon^a(\vec{x})$ depends on the space location and represents the amount of the transformation, and $T_a$ are the generators of the gauge transformations. 

The Lie algebra is given by
\begin{equation}
[T_a,T_b] = i f_{abc} T^c \,,
\end{equation}
where the $f_{abc}$ are the structure constants of the Lie group.  We can build from them the matrices $T^A$ that conform the adjoint representation of the Lie group:
\begin{equation}
(T^A_c)^a_{\;\;b} = -i f^a_{\;\;bc} \,.
\end{equation}

Once we define the induced metric on the group:
\begin{equation}
g_{ab} = Tr(T_a T_b) \,,
\end{equation}
which, by the way, can be written as\cite{weinberg}
\begin{equation}
g_{ab} = - f_{ac}^{\;\;\;\;d} f_{bd}^{\;\;\;\;c} \,,
\end{equation}
the following identities follow from the definitions above:
\begin{eqnarray}
Tr(T_a) &=& 0 \,, \\
Tr(T_a T_b T_c) &=& \frac{i}{2} f_{abc} + \frac{1}{2} Tr(T_a \{T_b,T_c\}) \,, \\
Tr(T_a T_b T_c T_d) &=& -\frac{1}{4} f_{ab}^{\;\;\;\;e} f_{cde} + \frac{i}{4} f_{cde} Tr(T_a \{T_b,T^e\}) \nonumber \\
&& + \frac{i}{4} f_{abe} Tr(T^e \{T_c,T_d\}) \nonumber \\
&& + \frac{1}{4} Tr(\{T_a, T_b\} \{T_c, T_d\}) \,,
\end{eqnarray}
and so on.

\subsection{The usual way}

I will first recall the usual arguments presented in textbooks, closely following Ref. \onlinecite{weinberg}.

As described above, a fermion field $\psi_l$ will transform under an infinitesimal local gauge transformation as:
\begin{equation}
\delta \psi_l = i g \epsilon^a (\vec{x}) (T_a)_l^{\;\;m} \psi_m \,.
\end{equation}

We want to employ not only $\psi_l$ but also $\partial_\mu \psi_l$ but, unfortunately, the latter does not transform as the fermion field:
\begin{equation}
\delta (\partial_\mu \psi_l) = ig [(\partial_\mu \epsilon^a(\vec{x})) (T_a)_l^{\;\;m} \psi_m + \epsilon^a(\vec{x}) (T_a)_l^{\;\;m} \partial_\mu \psi_m] \,,
\end{equation}
so we are in the necessity of building a covariant derivative of the fermion field, $D_\mu \psi_l$.  In order to do so, we must introduce a gauge field $A^a_\mu$:
\begin{equation}
D_\mu \psi_l = \partial_\mu \psi_ l - i g A^a_\mu (T_a)_l^{\;\;m} \psi_m \,,
\end{equation}
that transforms in the following way:
\begin{equation}
\delta A^a_\mu = \partial_\mu \epsilon^a(\vec{x}) + i \epsilon^b(\vec{x}) (T^A_b)^a_{\;\;c} A^c_\mu \,.
\end{equation}
Thus, the annoying $\partial_\mu \epsilon^a(\vec{x})$ term disappears when calculating $\delta (D_\mu \psi_l)$ making $D_\mu \psi_l$ transform as the fermion field.

The whole gauge-invariant Lagrangian can then be built from $\psi_l$ and $D_\mu \psi_l$, but not from $A^a_\mu$ (except when appearing in covariant derivatives).  However, we need to introduce a kinetic term for the vector field, so we have to consider terms of the form $\partial_\mu A^a_\nu$. This is accomplished by means of the gauge field strength tensor $F_{\mu \nu}^a$.  

The standard way of finding out $F_{\mu \nu}^a$ is described in Subsection \ref{1nonabe} and I will follow the most important steps here.  We will consider the commutator of two covariant derivatives acting on a fermion field:
\begin{equation}
[D_\mu, D_\nu] \psi_l = -i g (T_a)_l^{\;\;m} F^a_{\mu \nu} \psi_m \,, \label{strange}
\end{equation}
where the object $F^a_{\mu \nu}$ is defined as:
\begin{equation}
F^a_{\mu \nu} = \partial_\mu A^a_\nu - \partial_\nu A^a_\mu + g f^a_{\;\;bc} A^b_\mu A^c_\nu \,. \label{prifmn}
\end{equation}
Why is it so important to calculate $[D_\mu, D_\nu] \psi_l$ and to define $F^a_{\mu \nu}$?  Well, the point is that, since $[D_\mu, D_\nu] \psi_l$ transforms as a fermion field, we can calculate the transformation of $F^a_{\mu \nu}$, by means of Eq. (\ref{strange}), finding out that
\begin{equation}
\delta F^a_{\mu \nu} = i g \epsilon^b(\vec{x}) (T^A_b)^a_{\;\;c} F^c_{\mu \nu} \,,  \label{transfmn}
\end{equation}
i.e., $F^a_{\mu \nu}$ transforms in the adjoint representation of the group.  This object is then defined as the gauge field strength tensor.  Since the Lagrangian must contain a term quadratic in the ordinary derivatives of the gauge field, to build the kinetic term, it is reasonable to propose the Lorentz-invariant Lagrangian
\begin{equation}
\mathcal{L}_{K-A} = -\frac{1}{4} g_{ab} F^a_{\mu \nu} F^{\mu \nu b} \,,  \label{LKA}
\end{equation}
where $g_{ab}$ must be defined as
\begin{equation}
g_{ab} = Tr(T_a T_b) = - f_{ac}^{\;\;\;d} f_{bd}^{\;\;\;c} \,,
\end{equation}
so that the Lagrangian in Eq. (\ref{LKA}) is indeed gauge invariant.  $g_{ab}$ has the property of raising and lowering gauge indices and that is why it is called the induced metric on the group.  Thus, $\mathcal{L}_{K-A}$ becomes:
\begin{equation}
\mathcal{L}_{K-A} = -\frac{1}{4} F^a_{\mu \nu} F^{\mu \nu}_a \,.
\end{equation}

\subsection{The new way}
In the search of an intuitive, reasonable, natural, and perhaps more pedagogical way to find out $F^a_{\mu \nu}$, we can do better. 

Let's come back to the point just before Eq. (\ref{strange}).  In order to find out the appropriate form of the object resembling $\partial_\mu A^a_\nu$ we have to answer the question: how does this latter object transform under an infinitesimal gauge transformation?  Let's see it:
\begin{eqnarray}
\delta(\partial_\mu A^a_\nu) &=& \partial_\mu \partial_\nu \epsilon^a(\vec{x}) + i g (\partial_\mu \epsilon^b(\vec{x})) (T^A_b)^a_{\;\;c} A^c_\nu \nonumber \\
&& + i g \epsilon^b(\vec{x}) (T^A_b)^a_{\;\;c} \partial_\mu A^c_\nu \,.
\end{eqnarray}
The challenge is bigger than in the $\delta(\partial_\mu \psi_l)$ case since we have to get rid not only of the first-order derivatives of $\epsilon^a(\vec{x})$ but also of the second-order derivatives of it.  A clever way to get rid of the latter is to build an antisymmetrized object from $\partial_\mu A^a_\nu$:
\begin{eqnarray}
\delta(\partial_\mu A^a_\nu - \partial_\nu A^a_\mu) &=& i g (\partial_\mu \epsilon^b(\vec{x})) (T^A_b)^a_{\;\;c} A^c_\nu \nonumber \\
&& - i g (\partial_\nu \epsilon^b(\vec{x})) (T^A_b)^a_{\;\;c} A^c_\mu \nonumber \\
&& + i g \epsilon^b(\vec{x}) (T^A_b)^a_{\;\;c} (\partial_\mu A^c_\nu - \partial_\nu A^c_\mu) \,. \nonumber \\
&&
\end{eqnarray}
This stage seems to be a dead end since there does not seem to be a clear way of getting rid of the first-order derivatives of $\epsilon^a(\vec{x})$ unless the adjoint matrices vanish which is equivalent to have vanishing structure constants, i.e., to have an Abelian gauge theory.  Well... let's analyze this case.  

As said before, in an Abelian gauge theory, the structure constants vanish and, therefore, there does not exist an adjoint representation.  Therefore, we can use all the formulas already presented but taking into account that the generators $T_a$ are not matrices anymore whereas the matrices $T^A_a$ vanish. Thus, the fermion field transforms as
\begin{equation}
\delta \psi = i g \epsilon(\vec{x}) \psi \,,
\end{equation}
while the covariant derivative is defined as
\begin{equation}
D_\mu \psi = \partial_\mu \psi - i g A_\mu \psi \,,  \label{acd}
\end{equation}
where the gauge field must transform as
\begin{equation}
\delta A_\mu = \partial_\mu \epsilon(\vec{x}) \,,
\end{equation}
so that the covariant derivative does transform as the fermion field.  Thus, the object $\partial_\mu A_\nu$ transforms as
\begin{equation}
\delta (\partial_\mu A_\nu) = \partial_\mu \partial_\nu \epsilon(\vec{x}) \,,
\end{equation}
which makes the construction $\partial_\mu A_\nu - \partial_\nu A_\mu$ be the appropriate object since it is gauge invariant.  This object is called $F_{\mu \nu}$, the gauge field strength tensor, and since we require, in the Lagrangian, terms quadratic in the first-order derivatives of the gauge field, the following Lagrangian is obviously Lorentz invariant and gauge invariant:
\begin{equation}
\mathcal{L}_{K-A} = -\frac{1}{4} F_{\mu \nu} F^{\mu \nu} \,.
\end{equation}

An interesting aspect of the definition of $F_{\mu \nu}$ is that we can write
\begin{eqnarray}
F_{\mu \nu} &=& \partial_\mu A_\nu - i g A_\mu A_\nu - \partial_\nu A_\mu + i g A_\nu A_\mu \nonumber \\
&=& (\partial_\mu - i g A_\mu) A_\nu - (\partial_\nu - i g A_\nu) A_\mu \nonumber \\
&=& D_\mu A_\nu - D_\nu A_\mu \nonumber \\
&=& D_{[\mu} A_{\nu]} \,,
\end{eqnarray}
where the symbol $D_\mu$ is the same one as that defined in Eq. (\ref{acd}). $D_\mu A_\nu$ cannot be interpreted at all as the covariant derivative of the gauge field itself since such an object does not transform in the adjoint representation of the group which, in fact, does not exist:
\begin{equation}
\delta (D_\mu A_\nu) = \partial_\mu \partial_\nu \epsilon(\vec{x}) -i g (\partial_\mu \epsilon(\vec{x})) A_\nu - ig A_\mu \partial_\nu \epsilon(\vec{x}) \,.
\end{equation}

The expression $F_{\mu \nu} = D_{[\mu} A_{\nu]}$ should make us think that, maybe, the construction $D_{[\mu} A^a_{\nu]}$ is a good candidate to define the gauge field strength tensor in the non-Abelian case, the object built with first-order derivatives of the gauge field that transforms in the adjoint representation of the group.  Let's see if that is the case.

First of all, what is $D_\mu A^a_\nu$?  Let's remember that $D_\mu$ is a matrix:
\begin{equation}
D_\mu = \partial_\mu \mathbb{1} - ig A_\mu^b T_b \,,
\end{equation}
and so is $A_\nu$:
\begin{equation}
A_\nu = A^c_\nu T_c \,.
\end{equation}
Therefore, since we have
\begin{equation}
D_\mu A_\nu = (D_\mu A^a_\nu) T_a \,,
\end{equation}
we obtain
\begin{equation}
D_\mu A^a_\nu = Tr(D_\mu A_\nu T^a) \,.
\end{equation}
Thus, it is very easy to conclude that
\begin{equation}
D_\mu A^a_\nu = \partial_\mu A^a_\nu - ig A^b_\mu A^c_\nu \ Tr (T_b T_c T^a) \,.
\end{equation}
Such an expression allows us to build the object $D_{[\mu} A^a_{\nu]}$:
\begin{eqnarray}
D_{[\mu} A^a_{\nu]} &=& \partial_\mu A^a_\nu - \partial_\nu A^a_\mu \nonumber \\
&& - ig A^b_\mu A^c_\nu \ Tr (T_b T_c T^a) + ig A^b_\nu A^c_\mu \ Tr (T_b T_c T^a) \nonumber \\
&=& \partial_\mu A^a_\nu - \partial_\nu A^a_\mu \nonumber \\
&& - ig A^b_\mu A^c_\nu \ [Tr (T_b T_c T^a) - Tr (T_c T_b T^a)] \,,
\end{eqnarray}
which, by making use of the trace identities presented in Subsection \ref{prel}, reduces to
\begin{equation}
D_{[\mu} A^a_{\nu]} = \partial_\mu A^a_\nu - \partial_\nu A^a_\mu + g f^a_{\;\;bc} A^b_\mu A^c_\nu \,.
\end{equation}
This is precisely the gauge field strength tensor $F^a_{\mu \nu}$ found in Eq. (\ref{prifmn}) employing the usual methodology.  We already know from Eq. (\ref{transfmn}) that such an object transforms in the adjoint representation of the group but, in order to complete my whole argument and to make it completely logic and independent of the usual one, we can check by explicit calculations that $F^a_{\mu \nu} = D_{[\mu} A^a_{\nu]}$ does transform in the expected way.

\nocite{*}
%\bibliography{aipsamp}% Produces the bibliography via BibTeX.

\end{document}